\documentclass[twocolumn,floatfix]{revtex4-2}
\usepackage{amssymb}
\usepackage{natbib}
\usepackage{graphicx}
\usepackage{amsmath}
\usepackage[bookmarks = false,colorlinks = true,allcolors = black]{hyperref}
\usepackage{bm}
\usepackage[all]{hypcap}
\usepackage{graphicx}
\usepackage{colortbl}
\usepackage{booktabs}
\usepackage{enumerate}
\usepackage{enumitem}
\usepackage{braket}
\usepackage{mathrsfs}
\usepackage{multirow}
\usepackage{upgreek}
\usepackage{textcomp}

\usepackage{array}
\newcommand{\PreserveBackslash}[1]{\let\temp=\\#1\let\\=\temp}
\newcolumntype{C}[1]{>{\PreserveBackslash\centering}p{#1}}
\newcolumntype{R}[1]{>{\PreserveBackslash\raggedleft}p{#1}}
\newcolumntype{L}[1]{>{\PreserveBackslash\raggedright}p{#1}}

\begin{document}

\title{Experimental Generation of Spin-Photon Entanglement in Silicon Carbide}

\author{Ren-Zhou Fang$^{1,\,2,\,3}$}
\author{Xiao-Yi Lai$^{1,\,2,\,3}$}
\author{Tao Li$^{1,\,2,\,3}$}
\author{Ren-Zhu Su$^{1,\,2,\,3}$}
\author{Bo-Wei Lu$^{1,\,2,\,3}$}
\author{Chao-Wei Yang$^{1,\,2,\,3}$}
\author{Run-Ze Liu$^{1,\,2,\,3}$}
\author{Yu-Kun Qiao$^{1,\,2,\,3}$}
\author{Cheng Li$^{4}$}
\author{Zhi-Gang He$^{4}$}
\author{Jia Huang$^{5}$}
\author{Hao Li$^{5}$}
\author{Li-Xing You$^{5}$}
\author{Yong-Heng Huo$^{1,\,2,\,3}$}
\author{Xiao-Hui Bao$^{1,\,2,\,3}$}
\author{Jian-Wei Pan$^{1,\,2,\,3}$}
\affiliation{$^1$Hefei National Research Center for Physical Sciences at the Microscale and School of Physical Sciences, University of Science and Technology of China, Hefei 230026, China}
\affiliation{$^2$CAS Center for Excellence in Quantum Information and Quantum Physics, University of Science and Technology of China, Hefei, 230026, China}
\affiliation{$^3$Hefei National Laboratory, University of Science and Technology of China, Hefei 230088, China}
\affiliation{$^4$National Synchrotron Radiation Laboratory, University of Science and Technology of China, Hefei, Anhui 230029, China}
\affiliation{$^5$National Key Laboratory of Materials for Integrated Circuits, Shanghai Institute of Microsystem and Information Technology, Chinese Academy of Sciences, 865 Changning Road, Shanghai 200050, China}

\begin{abstract}
	A solid-state approach for quantum networks is advantages, as it allows the integration of nanophotonics to enhance the photon emission and the utilization of weakly coupled nuclear spins for long-lived storage. Silicon carbide, specifically point defects within it, shows great promise in this regard due to the easy of availability and well-established nanofabrication techniques. Despite of remarkable progresses made, achieving spin-photon entanglement remains a crucial aspect to be realized. In this paper, we experimentally generate entanglement between a silicon vacancy defect in silicon carbide and a scattered single photon in the zero-phonon line. The spin state is measured by detecting photons scattered in the phonon sideband. The photonic qubit is encoded in the time-bin degree-of-freedom and measured using an unbalanced Mach-Zehnder interferometer. Photonic correlations not only reveal the quality of the entanglement but also verify the deterministic nature of the entanglement creation process. By harnessing two pairs of such spin-photon entanglement, it becomes straightforward to entangle remote quantum nodes at long distance.
\end{abstract}

\maketitle

The hybrid entanglement between a matter qubit and a single photon serves as a fundamental resource for constructing quantum networks~\cite{KimbleKimble2008,WehnerHanson2018}, opening the door to remarkable applications such as distributed quantum computing and long-distance quantum communication through the use of quantum repeaters~\cite{BriegelZoller1998}. To enable the practical implementation of these applications, it is crucial that the matter-photon entanglement exhibits both high efficiency and long coherence times~\cite{SangouardGisin2011,WangPan2021}. The efficiency directly impacts the rate at which remote entanglement can be established, while the coherence time of the matter qubit determines the scalability of the system as the number of nodes increases. Solid-state systems~\cite{AwschalomZhou2018} offer several advantages in this regard. Nanoscale cavities~\cite{lodahl2015interfacing} can be built around matter qubits in order to significantly enhance the photon emission rate. Additionally, nuclear spins that couple weakly with environmental fields can be utilized for long-lived storage. 

One promising avenue for realizing these goals is the exploration of optically active defects. Extensive research has been conducted on color centers in diamond~\cite{robledo2011high,hensen2015loophole,RufHanson2021,pompili2021realization,hermans2022qubit,bhaskar2020experimental,stas2022robust}, leading to significant progress in demonstrating key components and functionalities for quantum networks. However, there is a growing desire to go beyond diamond and explore alternative host materials~\cite{AtatureWrachtrup2018,ZhangGali2020,WolfowiczAwschalom2021} that are more cost-effective and amenable to nanofabrication. In recent years, defects in silicon carbide~(SiC) have garnered significant interest~\cite{SonAwschalom2020}. Silicon carbide offers the advantages of large-scale, high-quality wafers that are already well-established in industry, as well as the feasibility of fabricating nanophotonic structures~\cite{LukinVuckovic2020}. Preliminary studies have successfully identified stable optical transitions~\cite{NagyWrachtrup2019,anderson2019electrical}, initiated and manipulated electron spins and nuclear spins~\cite{NagyWrachtrup2019,banks2019resonant,bourassa2020entanglement}, generated indistinguishable photons~\cite{MoriokaKaiser2020}, and demonstrated the feasibility of integration with nanophotonics~\cite{BabinWrachtrup2021,LukinVuckovic2020a,crook2020purcell,lukin2022optical}, long-lived storage and single-shot readout via charge state control~\cite{AndersonAwschalom2022}. However, the realization of the essential element of spin-photon entanglement still remains to be achieved.

\begin{figure*}[htb]
	\centering
	\includegraphics[width=1\textwidth]{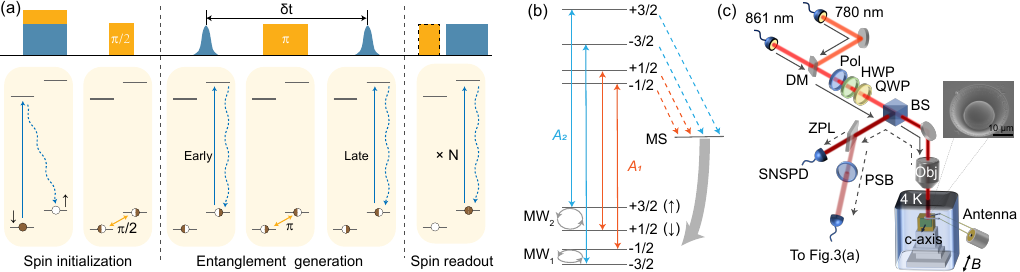}
	\caption{Experimental scheme and setup. (a) Experimental sequence consisting of spin initialization, entanglement generation and readout. (b) Energy level diagram of V1 centres. An external magnetic field parallel to the c-axis lift the degeneracy in ESs and GSs. Optical transitions $A_1$ and $A_2$, as well as spin transitions using MW$_1$, and MW$_2$, are present. We use $\ket{+3/2}_{g}$ as $\ket{\uparrow}$ state and $\ket{+1/2}_{g}$ as $\ket{\downarrow}$ state in the experimental protocol. Optical transitions with spin flip occur through the metastable state~(MS), which involves phonons and optical spontaneous emission. (c) Experimental setup. 
	BS: beamsplitter, Pol: polarizer, HWP: half-wave plate, QWP: 	quarter-wave plate.}
	\label{fig:1}
\end{figure*}

In this paper, we report the experimental generation of spin-photon entanglement in SiC. We make use of the V1 silicon vacancy~(V$_{\rm Si}$) in the epilayer of a 4H-SiC wafer that is created via high-energy electron irradiation~\cite{Supplement}. By applying a magnetic field in the $c$-axis, the V1 defect has an energy level diagram shown in Fig.~\ref{fig:1}b. With $S = 3/2$, both the ground and excited state have four sublevels, resulting in four optical transitions $\ket{m_s}_{g}\leftrightarrow\ket{m_s}_{e}$, with $m=$~$1/2$,~$-1/2$,~$3/2$,~$-3/2$. Since the g-factor is identical for the GSs and ESs, the four transitions are separated into two categories, with $A_1$ denoting the category $\ket{\pm1/2}_{g}\leftrightarrow\ket{\pm1/2}_{e}$, and $A_2$ denoting the category $\ket{\pm3/2}_{g}\leftrightarrow\ket{\pm3/2}_{e}$. The frequency difference between $A_1$ and $A_2$ is about 966~MHz due to the difference of zero-field splitting (ZFS) between the GSs and ESs. 

Our experimental scheme is shown in Fig.~\ref{fig:1}a. To form a qubit, we select two out of the four ground states sublevels, defining $\ket{\uparrow}=\ket{+3/2}_{g}$ and $\ket{\downarrow}=\ket{+1/2}_{g}$. We first initialize the defect to $\ket{\uparrow}$ and rotate it to a superpositional state of $(\ket{\uparrow}+\ket{\downarrow})/\sqrt{2}$ via applying a microwave $\pi/2$ pulse. Afterwards, we apply fast optical excitation for $\ket{\uparrow}$ and collect its photon emission. Then we add a microwave $\pi$ pulse and apply the fast optical excitation and emission collection again. A single-photon may be created either during the first or the second excitation step, thus generating a pair of spin-photon entanglement in the form of $\ket{\Psi}=\frac{1}{\sqrt{2}} (\ket{\uparrow}\ket{L} + \ket{\downarrow}\ket{E})$, where $\ket{E}$ denotes an early photon and $\ket{L}$ denotes a late photon. The correlations between the spin and photon states are measured in different basis by performing spin state projection and readout.

Our experimental setup is illustrated in Fig.~\ref{fig:1}c. The SiC sample is placed in a cryogen-free cryostat operated at 4 K. We make use the laser at 861~nm for resonant excitation and 780~nm for off-resonant excitation. The 861~nm laser is actively stabilized with an ultra-stable cavity to improve its phase coherence. The combined laser beams are focused on the sample with an objective~(Obj) len of NA = 0.65. Photon signal emitted from the sample is extracted through the reflection path of the $90:10$ BS. To improve the photon collection efficiency, we fabricate a solid immersion len~(SIL) around the defect via focused ion beam~(FIB) milling. Fluorescence at 861~nm is recognized as the zero-phonon line~(ZPL) emission, while fluorescence above 861~nm is recognized as the phonon sideband~(PSB) emission~\cite{Supplement}. In our experiment, we use a long-pass filter with an edge wavelength of 875~nm to split the ZPL and PSB. In addition we a bandpass filter~(855$\sim$865~nm) on the ZPL path to reject noise. 

\begin{figure}[htb]
	\centering
	\includegraphics[width=\columnwidth]{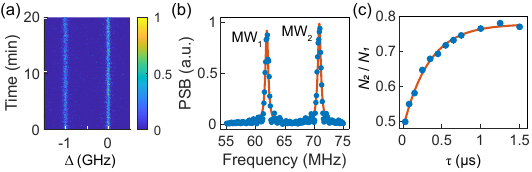}
	\caption{Optical and spin properties. (a) Resonant excitation scans over 20 min. The 780~nm laser depolarizes the GSs, then the PSB counts are collected in the following 2 $\upmu s$  resonant excitation. (b) ODMR signal after initialization into $\ket{\pm1/2}_{g}$ under 23.4 Gauss. Lines are fits using a Lorentzian function. (c) Radiative decay and spin-conserving transition ratio. After spin state initialization to $\ket{\uparrow}$, a two-pulse measurement with different intervals $\tau$ is used to obtain the ratio of radiative decay ratio and spin-conserving transition probability. The MS lifetime of 322 $\pm$ 24 ns is much longer than the excited state, so we approximate the ratio of fluorescence counts after the two pulse excitation ($N_2$ / $N_1$) at a $\tau =$ 30~ns as the proportion of radiative transitions, which is $50.0 \pm 0.6\%$. When $\tau$ is much longer than the MS lifetime, the proportion of returning to the $\ket{\pm 3/2}$ is $77.9\pm0.7\%$.}
	\label{fig:2}
\end{figure}

To implement the spin-photon entanglement scheme, a prerequisite is that the color center should exhibit stable and narrow optical transitions. We make use of resonant excitation and scan the laser frequency periodically and monitor the PSB signal meanwhile. The result is shown in Fig.~\ref{fig:2}a. It is clear to see that the transition frequency is stable during a measurement duration of 20~minutes. The excitation linewidth is estimated to be 73~MHz and 61~MHz for $A_1$ and $A_2$ optical transition, respectively, which are about 2.5 times the Fourier transform limit. By applying either $A_1$ or $A_2$, we can pump the defect to either $\ket{\pm3/2}_{g}$ or $\ket{\pm1/2}_{g}$, harnessing non-radiative decay channels via metastable doublet states shown Fig.~\ref{fig:1}b. The optical pumping fidelity is typically above 99\%.  In our experiment, two microwave transitions are employed for spin manipulation, namely MW$_1$ coupling $\ket{-1/2}_{g} \leftrightarrow \ket{-3/2}_{g}$ and MW$_2$ coupling $\ket{1/2}_{g} \leftrightarrow \ket{3/2}_{g}$. Starting from initial states $\ket{\pm 1/2}_{g}$, a typical optically detected magnetic resonance~(ODMR) result is shown in Fig.~\ref{fig:2}b by scanning the microwave frequency and monitoring the PSB signal. MW$_1$ is measured to be 62 MHz, and MW$_2$ is measured to be 70.8~MHz, with a separation of 8.8~MHz which corresponds to twice the ZFS for the GSs. 

\begin{figure}[htb]
	\centering
	\includegraphics[width=\columnwidth]{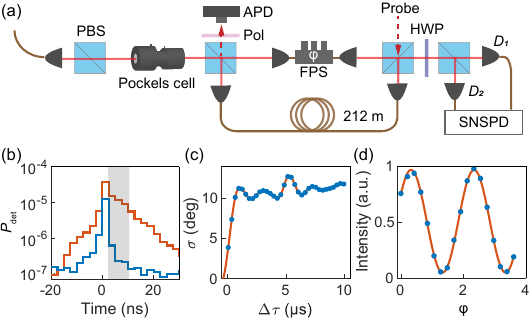}
	\caption{Measurement of the ZPL photon. (a) Unbalanced Mach-Zehnder interferometer for verification of entanglement in the time-bin degree.  (b) A histogram of the detection probability of ZPL and noise with the 10~ns detection window~(grey region). The orange line represents the total signal. The noise is quantified by utilizing a 5 GHz detuned laser~(blue line). $P_{det}$, detection probability. (c) Measurement of the laser phase noise at different time. The standard deviation $\sigma$ is 11$^\circ$ with $\Delta\tau$ = 1.06~$\upmu$s for this experiment. (d) Visibility fringes of the imbalanced interferometer.}
	\label{fig:3}
\end{figure}

The optical and spin characterizations of the defect provide a favorable starting point for implementing the spin-photon entanglement scheme. Further we need to extract the ZPL emission under resonant excitation. Selecting the ZPL emission is crucial for quantum network applications, since photons emitted from distant nodes needs to be highly indistinguishable to get remote nodes entangled. 
As the ZPL has the same frequency as the excitation beam, which eliminates the possibility of frequency filtering as used for PSB detection. In our experiment, we make use of polarization filtering and temporal filtering instead. We utilize a cross-polarization scheme~\cite{bernien2013heralded,nick2009spin}. The excitation and detection polarization are perpendicular to each other and both at $\theta$ = 45 $^\circ$ relative to the fluorescence polarization. Scattered light from the curved surface of SIL has varying polarization at different positions, and the vibration from the cold head leads to random drifting of the sample position, resulting in the extinction ratio dropping to 30~dB. The optical excitation $\pi$ pulse is made as short as 1 ns by using a fiber amplitude modulator (FAM) together with a home-built electrical fast-pulse generator. A typical ZPL signal is shown in Fig.~\ref{fig:3}b with a bin width of 2.5~ns. The excitation $\pi$ pulse opens in the $t=0$ time bin. We apply a temporal window from $t = 2.5$ to $t= 12.5$ to filter out the ZPL signal. The ZPL signal is over 30 times higher than the sum of the background contribution, including laser photons, dark count, and after-pulse from the superconducting nanowire single photon detector~(SNSPD).

To evaluate the spin-photon entanglement quality, we need to perform correlation measurements, which require measuring both the photon and the spin in various bases. To measure the time-bin encoded ZPL photon, we adopt an unbalanced Mach-Zehnder (MZ) interferometer, as showed in Fig.~\ref{fig:3}a. The Pockels cell directs the early photon to the long arm and late photon to the short arm, so the time-bin degree is converted to the polarization degree, i.e. $\ket{E}$ to $\ket{V}$ and $\ket{L}$ to $\ket{H}$. The relative time delay between the two arms matches the time difference of 1060~ns between the early and late mode to accommodate a $\pi$ MW$_2$ pulse of 920~ns. Preserving the relative phase $\varphi$ between the two optical modes is crucial for the measurements in the superpositional bases. Since the ZPL photon's phase follows with the excitation pulse. We first measure the phase noise of the 861~nm laser using another unbalanced MZ interferometer with 5~$\upmu$s delay~(not shown). The result shown in Fig.~\ref{fig:3}c. By calculating phase evolution $\varphi(\tau_0)-\varphi(\tau_0+\Delta\tau)$, we can deduce the the standard deviation of $\sigma$ being 11$^\circ$ with $\Delta\tau$ = 1.06~$\upmu$s. In addition, we need to keep the relative phase between two arms of the interferometer stable. In Fig.~\ref{fig:3}a, a probe beam with orthogonal polarization to the signal is introduced, following the same path as the signal. The phase information is converted into an intensity signal and detected by a sensitive avalanche photodiode (APD).  Relative phase variations are compensated for by the proportional-integral-derivative (PID) circuit and a fiber piezo stretcher (FPS) on the short arm. To test the interferometer's phase stability, we send a 1 ns resonant laser pulse with different phase setpoint. As shown in Fig.~\ref{fig:3}d, the interference visibility is 90\%, which leads to a reduction in the measured visibility of the superpositional basis. The imperfection of the visibility arises from the rapid fluctuations of the phase, which exceed the bandwidth of the FPS. 

The spin qubit is measured by detecting the PSB signal under resonant excitation. Under $A_2$ illumination, the PSB signal is proportional to the sum population in $\ket{\pm 3/2}_g$. Since our spin qubit is encoded in the subspace of 
$\ket{\uparrow}$ and $\ket{\downarrow}$, we can still use $A_2$ illumination to measure $\ket{\uparrow}$, and use MW$_2$ $\pi$ pulse following $A_2$ illumination to measure $\ket{\downarrow}$. The duration of this measurement is typically set to 1~$\upmu$s. While longer duration is unfavorable, as spin flip between $\ket{\pm 1/2}_g$ and $\ket{\pm 3/2}_g$ will happen and stop emitting photons. Measurement in a superpositional basis is performed by adding microwave rotations before the PSB detection. 

\begin{figure}[htb]
	\centering
	\includegraphics[width=\columnwidth]{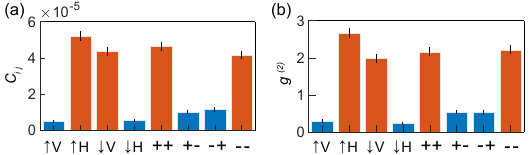}
	\caption{Coincidence and cross-correlations of the spin-photon entanglement. (a) Coincidence results. (b) Cross-correlations. Both measurements are performed in the eigenbasis and superpositional basis.}
	\label{fig:4}
\end{figure}

Finally we conduct the experiment of spin-photon entanglement generation with the control pulses set as Fig.~\ref{fig:1}a. We measure the coincidence counts on the eigenbasis and superpositional basis between the ZPL signal created in the entanglement generation phase and the PSB signal in the spin readout phase to characterize the entanglement quality. For measurements in the eigenbasis, the spin is projected to $\ket{\uparrow}$ or $\ket{\downarrow}$, the photon is projected to $\ket{H}$ or $\ket{V}$. While in the superpositional basis, the spin is projected to the states $\ket{\pm} = 1/\sqrt{2}(\ket{\uparrow} \pm \ket{\downarrow})$, and the photon is projected to $\ket{\pm} = 1/\sqrt{2}(\ket{H} \pm \ket{V})$. We define the the spin-photon coincidence probability as $C_{i\,j} = p_{i\,j}/\eta$ by normalizing the PBS detection probability, where $\eta$ denotes the PSB detection probability when the spin is fully in $\ket{\uparrow}$, $p_{i\,j}$ denotes the joint coincident probability of PSB and ZPL, with a subscript $i$ referring to the spin state and a subscript $j$ referring to the photon state. The results are given in Fig.~\ref{fig:4}a. To evaluate the entanglement, we first measure the visibility in the eigenbasis which is defined as 
\begin{equation}
V_e = \frac{C_{\uparrow H}+C_{\downarrow V}-C_{\uparrow V}-C_{\downarrow H}}{C_{\uparrow H}+C_{\downarrow V}+C_{\uparrow V}+C_{\downarrow H}},
\end{equation}
and get the result of $V_e = 81.0\% \pm 3.4\%$. Afterwards, we perform measurement in the superpositional basis. For the photonic state, this is achieved by rotating the half-waveplate in Fig.~\ref{fig:3}a. For the spin state, this is achieved via adding a $\pi/2$ or $3\pi/2$ pulse using MW$_2$. By using a similar definition, we get the superpositional visibility as $V_s =  60.9\% \pm 3.5\%$. From these two visibilities, we can estimate the entanglement fidelity as $F \approx (1 + V_e + 2V_s)/4 = 75.7\% \pm 1.5\%$. The result is significant higher than the bound of $50\%$ to certify entanglement. 

Multiple factors contribute to infidelity of the spin-photon entanglement. The dominant cause is the spin-flips during non-radiative decay. During the early excitation in Fig.~\ref{fig:1}a, the spin has a change of 50\% to decay through the MS levels in Fig.~\ref{fig:1}b. Further decay from the MS levels will result in spin flips. Since the interval of between the early excitation and the late excitation is much longer then the MS lifetime, the spin flips will result in some additional incoherent terms. This mechanism explains why we observe more coincidence counts for $C_{\uparrow H}$ than $C_{\downarrow V}$. These additional terms also explains the reduction of visibility in the superpositional basis partially since these terms are incoherent. For our current experiment, we estimate that this mechanism will result in a maximal fidelity of 90\%. We would like to note that this limitation will be mitigated significantly if we make use of the V2 defect instead. In this situation, the larger ZFS~(70~MHz) allows using a MW $\pi$ pulses that is much shorten than the MS lifetime. Accordingly to our estimation~\cite{Supplement}, a fidelity as high as 98.1\% can be achieved. 

In contrast to the entanglement generation process in an ensemble system~\cite{kuzmich2003generation}, the spin-photon entanglement generation process is in principle deterministic, albeit the emitted photon has a limited efficiency of lying in ZPL and being collected. This deterministic feature is verified via measuring the photonics cross correlation~$g^{(2)}$ between the ZPL and the PSB. Its definition is $g^{(2)}_{i\,j} = {p_{i\,j}}/{p_i \, p_j}$,
where $p_i$ denotes the PSB detection probability, $p_j$ denotes the ZPL detection probability. The measured results are shown in Fig.~\ref{fig:4}b. For a pair of ideal deterministic spin-photon entanglement, each set of measurement should has two terms with $g^{(2)} \simeq 2$ and two other terms with $g^{(2)} \simeq 0$. This feature is clearly verified by our results. We think slight deviation from the expected values is due to  instability of the charge state and the optical transition frequency, causing the V1 center to have a change to rest temporally and result in minor increase for $g^{(2)}$. 

The entanglement generation probability is established to be $2\times10^{-4}$~\cite{Supplement}, defined as the probability to get a ZPL photon in a SM fiber. The probability can be enhanced by incorporating the spin into nanocavities, which are coupled to on-chip waveguides~\cite{sipahigil2016integrated}. Efficient collection can be achieved with tapered-fiber waveguide coupling. As a 1D photonic crystal cavity with quality factor of $2\times10^4$ in this work~\cite{LukinVuckovic2020a}, the cavity coupling to single spin have a maximum Purcell factor of $F\sim$ 134. In this case, the Debye-Waller factor~($\eta_{d}$) will be improved from 8\% to 92\% and the quantum efficiency~($\eta_{q}$) will be improved from 50\% to 92\%. Taking into account the coupling loss to the feeding waveguide, the on-resonance transmission $\eta_t$ of the fundamental cavity mode can be achieved to 94\% according to this work~\cite{sipahigil2016integrated}. With the same manner of tapered-fiber waveguide coupling in this work~\cite{bhaskar2020experimental}, the coupling efficiency $\eta_c$ can be achieved to 93\%. The overall entanglement generation probability can be improved to $\eta = \eta_{d}\eta_{q}\eta_t\eta_c $ = 74\%.

In conclusion, we report the first experimental demonstration of spin-photon entanglement in silicon carbide, which is enabled by stable and narrow optical transitions, high-fidelity spin manipulation, and phase coherent dual-step optical excitations. With two pairs of such spin-photon entanglement it is straightforward to entangle two remote nodes via entanglement swapping~\cite{PanZeilinger1998,PanZukowski2012}. Fidelity of the spin-photon entanglement can be further improved significantly by making use of a MW $\pi$ pulse that is much shorter than lifetime of the metastable states. The entanglement generation rate can be further improved significantly via fabricating state-of-the-art photonic nanostructures around the defect~\cite{LukinVuckovic2020a,crook2020purcell,crook2020purcell} and harnessing high-efficient adiabatic coupling with a tapered fiber~\cite{sipahigil2016integrated}. For long-live storage, one can make use of the $^{29}$Si nuclear spin with recently demonstrated coherent control over individual nuclear spins and electron-nuclear spin pair with high fidelity~\cite{BabinWrachtrup2021}. By making the transition from electrical spin to nuclear, a long-lived spin-photon entanglement will be foreseeable in the very near future. By utilizing the nonlinearity of 4H-SiC, on-chip wavelength conversion to the telecom band can be implemented, which lead to ultra-compact chip-based integration.  Base on these improvements, the silicon vacancy defect in SiC may become a very promising approach for quantum networks. 

\begin{acknowledgments}
This research was supported by the Innovation Program for Quantum Science and Technology (No.~2021ZD 0301103), National Natural Science Foundation of China, and the Chinese Academy of Sciences.
\end{acknowledgments}

\setcounter{figure}{0}
\setcounter{table}{0}
\setcounter{equation}{0}

\onecolumngrid

\global\long\def\theequation{S\arabic{equation}}
\global\long\def\thefigure{S\arabic{figure}}
\global\long\def\thetable{S\arabic{table}}
\renewcommand{\arraystretch}{0.6}

\newpage

\newcommand{\msection}[1]{\vspace{\baselineskip}{\centering \textbf{#1}\\}\vspace{0.5\baselineskip}}

\msection{SUPPLEMENTAL MATERIAL}

\section{Sample preparation} 

The experiment sample is split from a 4-inch, 80~$\upmu$m thick epitaxial layer of single-crystal 4H-SiC grown on an n-type 4H-SiC wafer. V$_{\rm Si}$ centers are created by 10~MeV electron irradiation with a does of $2.3\times10^{12}$~cm$^{-2}$. After that, the sample was annealed at 500~$^{\circ}$C for 30~min to remove the interstitial defects. After etching markers with a spacing of 20~$\upmu$m on the side surface of the sample, We search for the V1 centers with good optical and spin coherence and record the position of them relative to the surrounding markers. Solid immersion lens (SIL), 15~$\upmu$m in diameter, are fabricated around the V1 centers to avoid total internal reflection of fluorescence. As shown in the insert picture of Fig. 1c, the deflection and convergence of light by the SIL result in a smaller offset of the focal point relative to the center of the SIL than the no SIL case.  After milling the SIL around the V1 center, we eliminate the bright emission layer near the surface by Reactive Ion Etching. To decrease Fresnel reflection we coat a single-layer anti-reflection film (aluminum oxide) on top of the sample surface resulting in a further enhanced collection efficiency. The V1 center used in this work is the brightest one among all the V1 centers in SILs.

\section{Optical and spin characterization}

\begin{figure*}[htb]
	\centering
	\includegraphics[width=\textwidth]{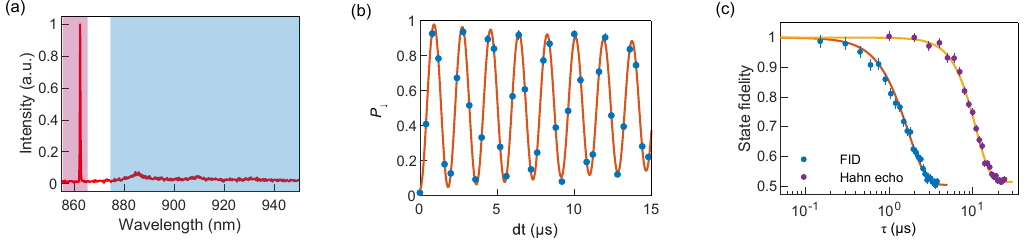}
	\caption{Optical and spin characterization. (a) Single V1 centre emission spectrum under off-resonant excitation.The pink region represents the wavelength collection range of the ZPL, from 855 to 865~nm. The blue region above 875~nm denotes the collection scope of the PSB. (b) Rabi oscillations for $\ket{\uparrow}\leftrightarrow\ket{\downarrow}$. Red lines are sinusoidal fits. (c) Free induction decay and Hahn echo measurement. In both measurements, the spin is initialized in $\ket{\uparrow}$. The state fidelity after free evolution time $\tau$ is measured by comparing with the state which experiences complete $\pi$ and 2$\pi$ MW$_2$ rotation in Ramsey and Hahn-echo measurement respectively. We infer $T_2^* = 1.68 \pm 0.05 \;\upmu s$ and $T_2 = 11.0\pm 0.2\;\upmu s$.}
	\label{fig:s1}
\end{figure*}

In our experiment, the ground state sublevels are manipulated with microwave pulses, which are coupled to the defect by shorting the inner conductor of a coaxial cable to the outer conductor with a 50-$\upmu$m-diameter silver wire. By applying microwave together with optical pumping, we can initialize the defect to a single sublevel within the ground state. For instance, with $A_1$ beam and MW$_1$ microwave, we can prepare an initial state of $\ket{\uparrow}$ with a fidelity around 96\%. Starting from $\ket{\uparrow}$, a typical result of Rabi oscillation is shown in Fig.~\ref{fig:s1}b. From this result, a microwave $\pi$ pulse fidelity is estimated to be around 96\%. In addition we have also performed a test of free induction decay with the result shown in Fig.~\ref{fig:s1}c, giving a dephasing time of $T_{2}^{*} = 1.68 \pm 0.05$~$\upmu$s. A spin echo extends the coherence time to 11~$\upmu$s further limited by rapid magnetic fluctuation from the paramagnetic spin bath in the sample.

\section{Infidelity due to non-radiative decay} 

The optical decay process through non-radiative channels after early excitation is the main source of the infidelity in our experimental principle. Here, we analyze the achievable fidelity by solely considering the non-radiative process. The density matrix of the generated entangled state is $\rho = \ket{\Psi}\bra{\Psi}+\alpha_1^2\ket{\uparrow,L}\bra{\uparrow,L}+\alpha_2^2\ket{-3/2,L}\bra{-3/2,L}$, where $\ket{\uparrow,L}\bra{\uparrow,L}$ and $\ket{-3/2,L}\bra{-3/2,L}$ are the incoherent terms. Based on the results shown in Fig.~2c, there is a probability of 28\% and 22\% for the spin decaying back to $\ket{\pm 3/2}$ and $\ket{\pm 1/2}$. Assuming the same decay rate from the MS state to two states in their subspace, if the non-radiative decay process ends before MW$_2$ $\pi$ pulse, $\alpha_1 = 0.055$ and $\alpha_2 = 0.07$. By renormalizing $\rho$, we estimate the fidelity is 90\% via $F=\mathrm{Tr}[(\ket{\Psi}\bra{\Psi})\rho]$. If we make use of the V2 defect instead, the microwave $\pi$ pulse can be shortened to $60$~ns. In this case, if non-radiative decay happens, the defect has a high chance to stay in the MS levels and does not contribute to photon generation during the late excitation. The proportion of the spin population decaying from MS state to GSs before late excitation in the total population decaying through non-radiative channels is 17\%. The coefficients are estimated to be $\alpha_1=9.35\times10^{-3}$ and $\alpha_2=1.19\times10^{-2}$, leading to a fidelity of 98.1\%.

\section{Entanglement generation probability} 

The entanglement generation probability is the probability of single ZPL photon in the SM fiber after a perfect optical $\pi$ excitation, which is related to the quantum efficiency, the Debye-Waller factor and the efficiency of collecting a ZPL photon into the SM fiber in the time-filtering window. The $\eta_{q}$ is 50\% as shown in Fig.~2c. The $\eta_d$ is about 8\%, which is determined as the fraction of the photon counts in the pink region in Fig.~\ref{fig:s1}a over the total counts under off-resonant excitation. The collection efficiency can be divided into four main parts: objective len, optical path, time-filtering and SM fiber. The transmittance of the objective len is 70\%. The SIL enhances the collection efficiency by a factor of seven compared to the bulk, resulting in 13\% of photon emission in the collection solid angle of the objective len. The total collection efficiency for the objective len is 9.1\%. The efficiency in the optical path is 33\%, which can be attribute to the 90:10 BS~(90\%), filters~(93\%) and the polarizer~(39\%). For the temporal filtering with a 6.7~ns lifetime of $A_2$ transition, the probability for the ZPL in the detection window shown in Fig.~3b is above 46\%. The collection efficiency into the SM fiber is about 36\%, estimated as the ratio of the photon counts collected into the SM fiber to those collected into the 50-$\upmu$m-core diameter multi-mode fiber. Taking into account all the factors mentioned above, the entanglement generation probability is $2\times10^{-4}$. The total efficiency of the unbalanced MZ interferometer, averaged over the short and long arms, is about 63\%. The quantum efficiency of the SNSPD is 85\%. Taking into account the losses introduced by the entanglement fidelity measurement devices, the detection probability for a ZPL photon is $1.1\times10^{-4}$, which is consistent with $p_j$.

\end{document}